\documentclass[info,debug,overfull]{epl}%
\usepackage{graphicx}
\usepackage{amsmath}
\usepackage{amsfonts}
\usepackage{amssymb}%
\institute{\inst{1}
Laboratoire de Physique de la Mati\`{e}re Condens\'{e}e, Coll\`{e}ge de France,
75231 Paris Cedex 05, France\\
\inst{2}
Department of Physics, Graduate School of Humanities and Sciences, Ochanomizu
University, 2--1--1, Otsuka, Bunkyo-ku, Tokyo 112-8610, Japan
}
\pacs{1}{64.70.Md Transitions in liquid crystals}
\pacs{2}{64.60.Qb Nucleation}
\pacs{3}{61.30.Vx Polymer liquid crystals}
\begin{document}

\title{Phase transitions of nematic rubbers}
\author{Pierre-Gilles de Gennes\inst{1} and Ko Okumura\inst{1,2}}
\maketitle

\begin{abstract}
Single crystal nematic elastomers undergo a transition from a strongly ordered
phase $N$ to an "isotropic" phase $I$. We show that: (a) samples produced
under tension by the Finkelmann procedure are intrinsically anisotropic and
should show a small (temperature dependent) birefringence in the high
temperature $I$ phase. (b) for the $I\rightarrow N$ transition \textit{via}
cooling there is a spinodal limit but for the $N\rightarrow I$ transition
\textit{via} heating there is no soft mode at the standard spinodal
temperature. (c) the $N\rightarrow I$ transition is reminiscent of a
\textit{martensitic transformation}: nucleation of the $I$ phase should occur
in the form of platelets, making a well defined angle with the director.

\end{abstract}

\shortauthor{P.-G. de Gennes and K. Okumura}

\section{Introduction}

Nematic rubbers were first constructed in a single domain form by H.
Finkelmann and coworkers \cite{1}. They show a spectacular change in shape
when they are switched from low temperatures ($T<T_{c}$) to high temperatures
($T>T_{c}$). Many properties resulting from the coupling between nematic order
and elastic deformations \cite{2} have been analyzed by M. Warner, E. M.
Terentjev and coworkers \cite{3}. We are concerned here mainly by the
transition from the nematic phase $N$, to the high temperature phase $I$.

Two striking facts should be mentioned here: (a) the transition is expected to
be first order, but the plots of birefringence versus temperature $T$ are
continuous. \cite{6} (b) the transition is very slow (minutes). \cite{WT}

Our aim here is to discuss some effects of the nematic / elastic coupling on
the phase transition. First, we discuss the anisotropy induced by the
preparation method. We show that this preparation imposes a fixed
\textit{external field} coupled to the order parameter: under this field the
plots of order versus temperature are expected to be continuous. Second, we
analyze $T$-jumps: for a cooling jump we find a traditional spinodal
transition. But for a heating jump we show that the elastic couplings tend to
suppress the spinodal instability. Finally, we discuss the nucleation of an
$I$ phase inside a nematic single crystal, and show that it should take place
in the form of platelets, with well defined geometrical conditions.

\section{Anisotropy effects}

In the Finkelmann scheme, a very weak network is prepared first, and is put
under a prescribed deformation $\varepsilon_{zz}^{0}$ along one axis. \cite{1}
Then a second reaction is started and the final nematic rubber is generated.
The basic energy for this situation is of the form%
\begin{equation}
F=F_{0}\left(  Q_{\alpha\beta}\right)  -\Lambda Q_{\alpha\beta}\varepsilon
_{\beta\alpha}+\mu\left(  \varepsilon_{\alpha\beta}-\varepsilon_{\alpha\beta
}^{0}\right)  \left(  \varepsilon_{\beta\alpha}-\varepsilon_{\beta\alpha}%
^{0}\right)  \label{e1}%
\end{equation}
where summation over the repeated indices is implied. For simplicity, we
restrict our attention to small deformations $\varepsilon_{\alpha\beta}$.
$F_{0}$ is the standard Landau free energy of the nematic order, $\Lambda$
describes the coupling between deformation and order, and the last term is the
shear elastic energy (we consider only incompressible systems: $Q_{\alpha
\beta}$ and $\varepsilon_{\alpha\beta}$ are \textit{symmetric traceless
tensors}). The crucial point in Eq. (\ref{e1}) is that the elastic energy
(with a strong coefficient $\mu$) is minimal \textit{in the original state}
($\varepsilon=\varepsilon^{0}$). We can shift the definition of deformations,
writing%
\begin{equation}
\varepsilon=\varepsilon^{0}+e
\end{equation}
where $e$ measures actual deformations from the high temperature $I$ phase.
Then the energy is (in contracted notation)%
\begin{equation}
F=F_{0}-\Lambda Q\varepsilon_{0}-\Lambda Qe+\mu e^{2}%
\end{equation}
There is a \textit{constant field} $\sigma_{0}=\Lambda\varepsilon_{0}$ acting
on the order parameter. This linear term in $Q$ implies that, in the nominally
isotropic $I$ phase, we have a non-zero order $Q$.

Writing explicitly for small $Q$%
\begin{equation}
F=A_{0}(T)Q_{\alpha\beta}Q_{\beta\alpha}-\sigma_{\alpha\beta}^{0}%
Q_{\beta\alpha}-\Lambda Q_{\alpha\beta}e_{\beta\alpha}+\mu e_{\alpha\beta
}e_{\beta\alpha}%
\end{equation}
\textit{Via} minimization in terms of $e_{\alpha\beta}$:%
\begin{equation}
e_{\alpha\beta}=\frac{\Lambda}{2\mu}Q_{\alpha\beta} \label{eQ}%
\end{equation}
we first arrive at the classical renormalization of $A_{0}(T)\simeq
a_{0}\left(  T-T_{0}\right)  $:%
\begin{equation}
F=a(T)Q_{\alpha\beta}Q_{\beta\alpha}-\sigma_{\alpha\beta}^{0}Q_{\beta\alpha}
\label{Fm}%
\end{equation}
where%
\begin{equation}
a(T)=A_{0}(T)-\frac{\Lambda^{2}}{4\mu}\equiv A_{0}(T)-a_{0}\Delta T_{c}
\label{Tc}%
\end{equation}
This minimized energy gives a finite high temperature order:%
\begin{equation}
Q_{\alpha\beta}^{I}=\frac{\sigma_{\alpha\beta}^{0}}{2a(T)}%
\end{equation}
This describes a birefringence which is high near the transition point (where
$a$ is small) very much like the Kerr effect in standard nematics. This may be
an explanation for the results of ref. \cite{6}.

In the following, we employ a Landau expansion in terms of $Q\rightarrow
Q-Q^{I}$, and set the free energy to be zero at this "isotropic" phase, which
now corresponds to $Q=0$.

\section{Absence of spinodal instabilities}

Can we achieve the transition by soft phonon modes? Since our materials are
essentially incompressible, we must investigate the transverse phonons. We
assume rapid equilibration for the elastic degrees of freedom: we always use
Eq. (\ref{eQ}).

For a transverse phonon of wave vector $%
%TCIMACRO{\TeXButton{vectorQ}{\mbox{ \boldmath$q$ }}}%
%BeginExpansion
\mbox{ \boldmath$q$ }%
%EndExpansion
$ and displacement field $%
%TCIMACRO{\TeXButton{vectorU}{\mbox{ \boldmath$u$ }}}%
%BeginExpansion
\mbox{ \boldmath$u$ }%
%EndExpansion
$ (with $%
%TCIMACRO{\TeXButton{vectorQ}{\mbox{ \boldmath$q$ }}}%
%BeginExpansion
\mbox{ \boldmath$q$ }%
%EndExpansion
\cdot%
%TCIMACRO{\TeXButton{vectorU}{\mbox{ \boldmath$u$ }}}%
%BeginExpansion
\mbox{ \boldmath$u$ }%
%EndExpansion
=0$) the deformation is%
\begin{equation}
e_{\alpha\beta}=\frac{1}{2}\left(  q_{\alpha}u_{\beta}+q_{\beta}u_{\alpha
}\right)
\end{equation}

a) Consider first the instability of the disordered phase upon
\textit{cooling} ($I\rightarrow N$ transition). Here the Landau free energy in
Eq. (\ref{Fm}) for small but nonzero amplitudes reduces to
\begin{equation}
F=a(T)Q_{\alpha\beta}Q_{\beta\alpha}%
\end{equation}
and at the spinodal temperature $T^{\ast\ast}$ the coefficient $a(T^{\ast\ast
})$ vanishes. Note again that in this expansion, $Q$ is actually the
difference from $Q^{I}$.

Amplitudes of our transverse phonons are proportional to $Q_{\alpha\beta}$
(Eq. (\ref{eQ})) and the energy required for the phonon generation vanishes at
$T=T^{\ast\ast}$: the phonons are soft at this temperature.

b) the situation is different for the $N\rightarrow I$ transition upon
\textit{heating}. Here we start with a uniaxial nematic phase with nonzero
components $(Q_{xx},$ $Q_{yy},$ $Q_{zz})=(-S/2,$ $-S/2,$ $S)$; we can find a
candidate for the spinodal temperature $T^{\ast}$ for a special value
$S=S^{\ast}$, satisfying the conditions%
\begin{equation}
\frac{\partial F}{\partial S}=\frac{\partial^{2}F}{\partial S^{2}}=0\text{
\ (at }S=S^{\ast}\text{ and }T=T^{\ast}\text{)} \label{spinodal}%
\end{equation}
We employ the Landau expansion (which is already minimized for the elasticity)%
\begin{equation}
F-F_{0}=aQ_{\alpha\beta}Q_{\beta\alpha}-bQ_{\alpha\beta}Q_{\beta\gamma
}Q_{\gamma\alpha}+c\left(  Q_{\alpha\beta}Q_{\beta\alpha}\right)  \left(
Q_{\gamma\delta}Q_{\delta\gamma}\right)
\end{equation}
where $b$ and $c$ are positive (see Ch. 2 of ref. \cite{rem}).

We expand the free energy to second order around the spinodal point. At this
point the coefficient $\partial^{2}F/\partial Q_{zz}^{2}=0$ but the other
curvature coefficients do not vanish. We find:%
\begin{equation}
F(Q)-F(Q^{\ast})=\frac{9b^{2}}{64c}\left[  \left(  Q_{xx}-Q_{yy}\right)
^{2}+4Q_{xy}^{2}\right]  \label{fexpand}%
\end{equation}
The linear term in $\Delta Q_{\alpha\beta}\equiv Q_{\alpha\beta}%
-Q_{\alpha\beta}^{\ast}$ and the $\Delta Q_{zz}^{2}$ term vanish as expected
from Eq. (\ref{spinodal}). In addition, there is no term proportional to
$Q_{zx}^{2}$ and $Q_{zy}^{2}$: they are ruled out by the rotational invariance
around the director axis $z$.

\begin{figure}[h]
\includegraphics[height=2.5cm]{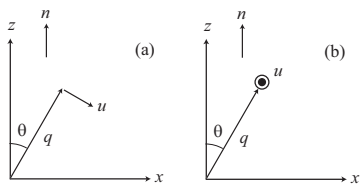}\caption{{}(a) First transverse
mode$\mbox{ \boldmath$u$ }=(u\cos\theta,0,-u\sin\theta)$. (b) Second
transverse mode$\mbox{ \boldmath$u$ }=(0,u,0)$. Note the difference from Fig.
\ref{f1}: the director vector is parallel to the $z$-axis.}%
\label{fa1}%
\end{figure}

We can now investigate the transverse phonon modes (Fig. \ref{fa1}).

The first mode, with$%
%TCIMACRO{\TeXButton{vectorQ}{\mbox{ \boldmath$q$ }}}%
%BeginExpansion
\mbox{ \boldmath$q$ }%
%EndExpansion
$and$%
%TCIMACRO{\TeXButton{vectorU}{\mbox{ \boldmath$u$ }}}%
%BeginExpansion
\mbox{ \boldmath$u$ }%
%EndExpansion
$in the $x-z$ plane (Fig. \ref{fa1}a), induces a component proportional to
$Q_{xx}$ ($\simeq q_{x}u_{x}=\varepsilon\sin\theta\cos\theta$ with
$\varepsilon\equiv qu$), but not to $Q_{yy}$. Thus, it contributes to the
$\left(  Q_{xx}-Q_{yy}\right)  ^{2}$ term in the free energy, which is
positive ($\simeq\mu e_{xy}^{2}\sim\left(  \varepsilon\sin\theta\right)  ^{2}%
$). A finite $Q_{xx}-Q_{yy}$ implies a biaxial nematic, and this costs
energy.

The second mode, with$%
%TCIMACRO{\TeXButton{vectorU}{\mbox{ \boldmath$u$ }}}%
%BeginExpansion
\mbox{ \boldmath$u$ }%
%EndExpansion
$normal to both$%
%TCIMACRO{\TeXButton{vectorQ}{\mbox{ \boldmath$q$ }}}%
%BeginExpansion
\mbox{ \boldmath$q$ }%
%EndExpansion
$and director vector, induces an amplitude of $Q_{xy}$ ($\simeq q_{x}%
u_{y}=\varepsilon\sin\theta$): again this induces biaxiality and requires a
finite energy ($\simeq\mu e_{xy}^{2}\sim\left(  \varepsilon\sin\theta\right)
^{2}$).

We conclude that both transverse modes are not soft for all nonzero angles
($\theta\neq0$) at $T=T^{\ast}$. The $\theta=0$ mode is soft at all
temperatures, if $F(Q)$ has full rotational invariance. But this mode cannot
cause a change in the magnitude of the order parameter and does not catalyze
the transition.

\section{Nucleation of a high temperature $I$ phase}

We consider now a $T$-jump from a temperature just below the thermodynamic
transition point $T_{c}$ towards a higher temperature $T$, in the region where
spinodal instabilities are ruled out.

\noindent\emph{A) Choice of a habit plane}

We now investigate a possible plane boundary (\textit{habit plane}) between
$I$ and $N$ phase.

\noindent\emph{A.1) The two-dimensional case (plane strain)}%

%TCIMACRO{\TeXButton{Fig2-3}{\begin{figure}[!]
%\twofigures[height=5.5cm]{habit.eps}{disc.eps}
%\caption{(a) Habit plane separating nematic and isotropic regions. In two
%dimensions the "isotropic" network is represented by a square array, while
%the nematic network is represented by parallelograms. Along the habit plane
%the two networks match. (b) Detailed definitions of angles and axes.}
%\label{f1}
%\caption{Platelet nucleus in the nematic phase accompanied by a large
%strain around the periphery.}
%\label{f4}
%\end{figure}}}%
%BeginExpansion
\begin{figure}[!]
\twofigures[height=5.5cm]{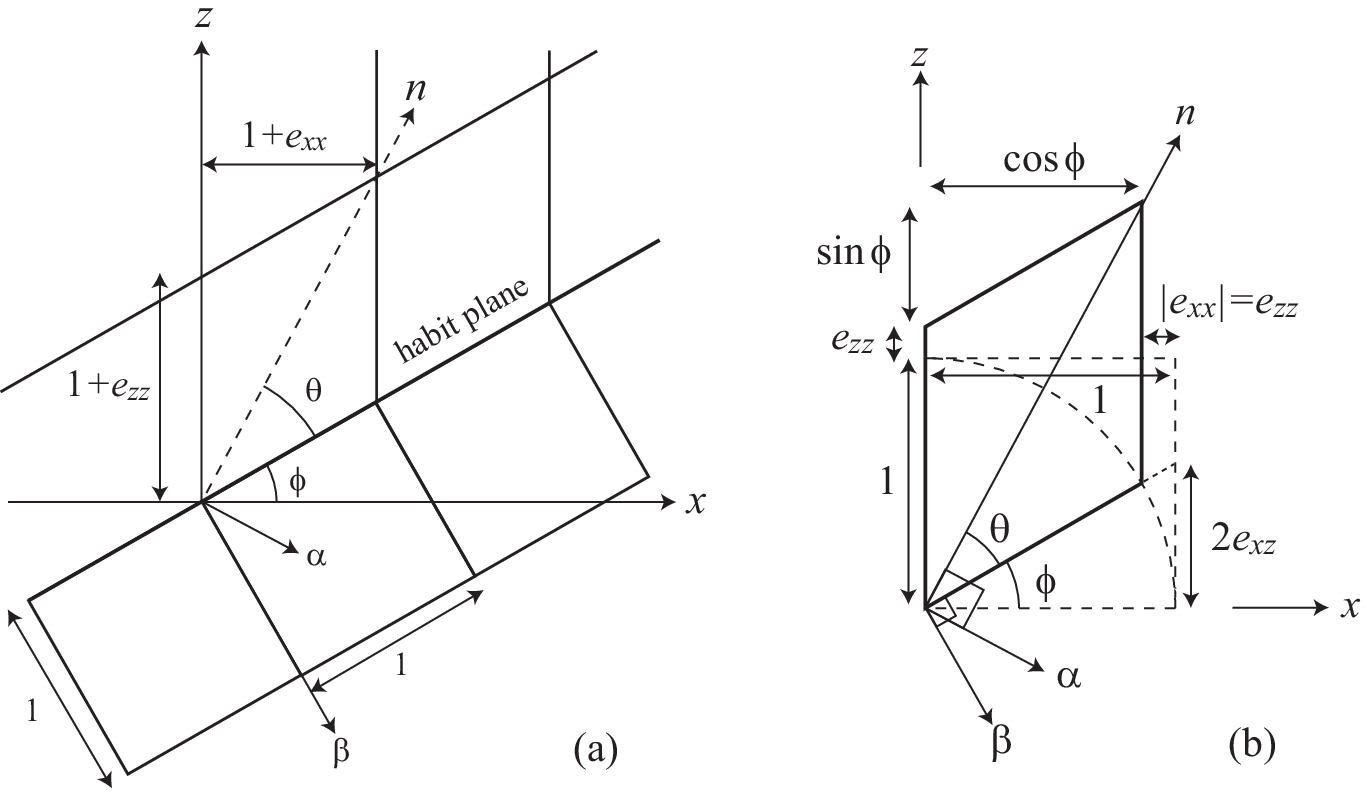}{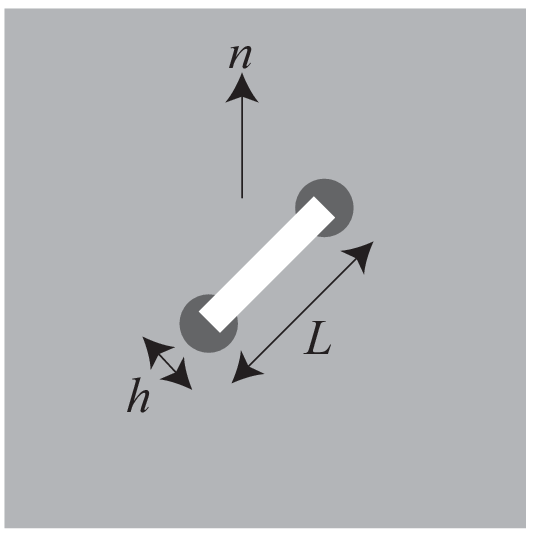}
\caption{(a) Habit plane separating nematic and isotropic regions. In two
dimensions the "isotropic" network is represented by a square array, while
the nematic network is represented by parallelograms. Along the habit plane
the two networks match. (b) Detailed definitions of angles and axes.}
\label{f1}
\caption{Platelet nucleus in the nematic phase accompanied by a large
strain around the periphery.}
\label{f4}
\end{figure}%
%EndExpansion

This two-dimensional example (in the $x-z$ plane) is shown on Fig. \ref{f1},
where we impose compatibility between an $I$ phase (represented symbolically
by a square unit cell) and a uniaxial $N$ phase (represented by
parallelograms). The squares have been rotated by an angle $\phi$ from the
$x$-axis. The $N$ phase has an elongation $e_{zz}$ along the $z$ axis, which
will be determined below. Note here that the director is \textit{not} parallel
to the $z$ axis; the parallelograms are sheared with a strain%
\begin{equation}
e_{xz}=\frac{1}{2}(\partial_{x}u_{z}+\partial_{z}u_{x})=\frac{1}{2}\tan\phi
\end{equation}
Here and hereafter displacement fields are defined in the $x-z$ frame shown in
Fig. \ref{f1}b. There is also a deformation $e_{xx}=-e_{zz}$ in the $N$ phase
(we assume incompressibility) and matching imposes $1+e_{xx}=\cos\phi$. The
two-dimensional deformation matrix is written as%
\begin{equation}
\widehat{e}\equiv\left(
\begin{array}
[c]{cc}%
e_{xx} & e_{xz}\\
e_{xz} & e_{zz}%
\end{array}
\right)  =\left(
\begin{array}
[c]{cc}%
\cos\phi-1 & \frac{1}{2}\tan\phi\\
\frac{1}{2}\tan\phi & 1-\cos\phi
\end{array}
\right)  . \label{eq2}%
\end{equation}
Thus from Fig. \ref{f1}b the angle $\theta$ between the director axis and the
habit plane is given by%
\begin{equation}
\tan\left(  \theta+\phi\right)  =\frac{1+e_{zz}+\sin\phi}{\cos\phi}%
=\frac{2-\cos\phi+\sin\phi}{\cos\phi}%
\end{equation}
Note that, for small $\phi$, the parallelogram is a lozenge, and the director
is parallel to the long axis of the lozenge -- at $\pi/4$ from the habit plane
(Fig. \ref{f4} corresponds to this limit).

The magnitude of $\phi$ is fixed by the condition that the elastic energy of
$N$ state be equal to the equilibrium value $f(T_{i})$ in the initial nematic
phase. Under the incompressibility condition, we have $f(T_{i})=\mu(e_{xx}%
^{2}+e_{zz}^{2}+2e_{xz}^{2})=2\mu(e_{xx}^{2}+e_{xz}^{2})$. Thus, $\phi$ is
determined by%
\begin{equation}
f(T_{i})/(2\mu)=(1-\cos\phi)^{2}+\frac{1}{4}\left(  \tan\phi\right)  ^{2}%
\end{equation}
For this two-dimensional case, we achieve a perfectly isotropic state $I$ on
one side and the standard $N$ state of the other side. However, as we shall
see, this perfect matching is not possible in 3D (exactly as in similar
martensitic transformations \cite{7}).

\noindent\emph{A.2) The three-dimensional case}

We can try a similar construction for the three dimensional case (see Fig.
\ref{f1} again). We take the habit plane perpendicular to the $x-z$ plane (the
director axis stays in the $x-z$ plane). In the rectangular coordinate
composed of the director axis ($n$), $y$ axis, and $\alpha$ axis
(perpendicular to the others, see Fig. \ref{f1}), the non-zero strain
components are $e_{n}(T_{i})\equiv e_{i}\neq0$ (in the director axis), and
$e_{\alpha\alpha}=e_{yy}=-e_{i}/2$. The elastic energy associated with this is
given by $f_{i}=\frac{3}{2}\mu e_{i}^{2}$. In the $(x,y,z)$ coordinate the
strain tensor is given as%
\begin{equation}
\widehat{e}\equiv\left(
\begin{array}
[c]{ccc}%
e_{xx} & e_{xy} & e_{xz}\\
e_{yx} & e_{yy} & e_{yz}\\
e_{zx} & e_{zy} & e_{zz}%
\end{array}
\right)  =\left(
\begin{array}
[c]{ccc}%
\cos\phi-1 & 0 & \frac{1}{2}\tan\phi\\
0 & -e_{i}/2 & 0\\
\frac{1}{2}\tan\phi & 0 & 1-\cos\phi+e_{i}/2
\end{array}
\right)
\end{equation}
and the elastic energy is given by $f_{i}/\mu=e_{xx}^{2}+e_{yy}^{2}+e_{zz}%
^{2}+2e_{zx}^{2}$. Thus, we can fix $\phi$ by requiring that the energy is the
same in the $(n,y,\alpha)$ and $(x,y,z)$ coordinates:%
\begin{equation}
\frac{5}{4}e_{i}^{2}=\left(  \cos\phi-1\right)  ^{2}+\left(  1-\cos\phi
+e_{i}/2\right)  ^{2}+\frac{1}{2}\left(  \tan\phi\right)  ^{2} \label{eqa1}%
\end{equation}
In addition, we have%
\begin{equation}
\tan\left(  \theta+\phi\right)  =\frac{2-\cos\phi+e_{i}/2+\sin\phi}{\cos\phi}
\label{eqa2}%
\end{equation}
We note that Eqs. (\ref{eqa1}) and (\ref{eqa2}) predict that the habit angle
approaches $\pi/4$ again in the small deformation limit ($e_{i},\phi
\rightarrow0$).

In this three-dimensional case, the $N$ phase has a third component of strain
$e_{yy}=-e_{i}/2$ and this can not be eliminated in the $I$ phase. From the
incompressibility condition in the $I$ phase, we must also have a strain
$e_{\beta\beta}=-e_{yy}$; inside the nucleus, there must be non-zero strain
components $(e_{yy},e_{\beta\beta})=(-e_{i}/2,e_{i}/2)$, which results in a
residual elastic energy (per unit volume)
\begin{equation}
E_{I}=\mu e_{i}^{2}/2.
\end{equation}
Thus we reach a state which is not totally isotropic

\noindent\emph{B) Homogeneous nucleation}

Again we think of a $T$-jump from an initial temperature $T_{i}$ just below
$T_{c}$ (we take $T_{i}=T_{c}$ in practice for simplicity) up to a final
temperature $T$ lying about 10 degrees above $T_{c}$. Our assumed shape for
the nucleus is a platelet corresponding to the habit plane orientation, with a
thickness $h$ and dimension $L$ (Fig. \ref{f4}). From now on we construct only
crude estimates, ignoring all numerical coefficients: for instance the
platelet volume is taken to be $L^{2}h$.

The platelet energy is of the form%
\begin{equation}
f=-L^{2}h\cdot\Delta+E_{I}L^{2}h+\gamma L^{2}+E_{p} \label{eq5}%
\end{equation}
Here $\Delta$ is the gain in free energy (per unit volume) obtained by
switching from $N$ to $I$ at the temperature $T$. Thermodynamics imposes%
\begin{equation}
\Delta=\left(  T-T_{c}\right)  \Delta S \label{del}%
\end{equation}
where $\Delta S\simeq a_{0}Q^{2}$ is the entropy jump at the transition
temperature $T_{c}$.

The second term $E_{I}$ in Eq. (\ref{eq5}) is due to the residual stress in
the $I$ phase [The principle axes of the deformation here are the $y$ and
$\beta$ axes and another axes normal to them, i.e. the non-zero elements of
the diagonalized deformation matrix are $(e_{yy},e_{\beta\beta})=(-e_{i}%
/2,e_{i}/2)$] and is only a fraction of the initial elastic energy. The
original elastic energy is $\frac{3}{2}\mu e_{i}^{2}$ while we have here
$\mu(e_{yy}^{2}+e_{\beta\beta}^{2})=\frac{1}{2}\mu e_{i}^{2}$.

Thus we lump the elastic correction into $\Delta$, writing%
\begin{equation}
\widetilde{\Delta}=\Delta-\frac{1}{2}\mu e_{i}^{2}\equiv\left(  T-T_{m}%
\right)  \Delta S \label{delb}%
\end{equation}
where we have introduced $T_{m}$ in analogy with Eq. (\ref{del}). $T_{m}$ is
the minimal final temperature for nucleation. From a scaling point of view, we
should have%
\begin{equation}
\frac{\widetilde{\Delta}}{\mu e_{i}^{2}}\simeq\frac{T-T_{m}}{\Delta T_{c}}
\label{ratio}%
\end{equation}
where $\Delta T_{c}$ corresponds to an increase in the $N-I$ transition
temperature due to the coupling (See Eq. (\ref{Tc})). This can be explicitly
shown by using the relation $\mu e_{i}^{2}\simeq\Lambda^{2}Q^{2}/\mu$ and
$\Delta S\simeq a_{0}Q^{2}$. We assume from now on that $\widetilde{\Delta}$
is positive but small.

The third term in Eq. (\ref{eq5}) is a surface term, with a certain
interfacial energy $\gamma$, which might be estimated by a Ginzburg - Landau
type free energy (including the spatial dependence of the order parameter)
plus elastic tensors. Because of these latter terms $\gamma$ is positive even
when $T>T^{\ast}$ as indicated by the previous arguments on the absence of the
spinodal instability.

The last term in Eq. (\ref{eq5}) is another elastic term, due to distortions
at the \textit{periphery} of our platelet. Here, there is no matching at all
(the matching was achieved only at the habit plane boundary), we have
\emph{large deformations} taking place in a toroidal region near the periphery
(Fig. \ref{f4}). The length scale relevant for this extra peripheral
deformation and the size of deformation are both around $h$ and, because the
deformation fields always satisfies the Laplace equation, this deformation
dies out only at a distance $\simeq h$ from the torus: the strain $e_{p}$ (at
least of the order of $e_{i}$) is stored around the periphery of the volume
$\simeq Lh^{2}$ (Fig. \ref{f4}):%
\begin{equation}
E_{p}\simeq Lh^{2}\Delta_{p}%
\end{equation}
with%
\begin{equation}
\Delta_{p}=\mu e_{p}^{2}%
\end{equation}

We now write Eq. (\ref{eq5}) in the following form:%
\begin{equation}
F=-\widetilde{\Delta}L^{2}h+\gamma L^{2}+Lh^{2}\Delta_{p}%
\end{equation}

By minimizing the above energy with respect to $h$, we have%
\begin{align}
h_{0}  &  =\frac{\widetilde{\Delta}}{2\Delta_{p}}L\\
F_{0}  &  =-\frac{\widetilde{\Delta}}{4\Delta_{p}}\widetilde{\Delta}%
L^{3}+\gamma L^{2} \label{F0}%
\end{align}
Thus the assumption $L\gg h$ corresponds to $\widetilde{\Delta}\ll\Delta_{p}$:
if the peripheral energy $\Delta_{p}$ is large compared with the bulk gain
$\widetilde{\Delta}$, the platelet tends to become thin.

The energy expression in Eq. (\ref{F0}) is in contrast with the classical case
of a spherical nucleus without any elastic contribution: $F=-\Delta\cdot
L^{3}+\gamma L^{2}$. For $\Delta_{p}>\widetilde{\Delta}$, the bulk term for
the disk droplet is quite small compared with the surface term. If we look for
the maximum of Eq. (\ref{F0}) with respect to $L$, we find an energy barrier
$U=F_{0}(L^{\ast})$ for the formation of a critical droplet:%
\begin{align}
L^{\ast}  &  =\frac{\gamma}{\widetilde{\Delta}}\cdot\frac{8\Delta_{p}%
}{3\widetilde{\Delta}}\\
U  &  =\frac{2}{3}\left(  \frac{8\Delta_{p}\Delta}{3\widetilde{\Delta}^{2}%
}\right)  ^{2}U_{c} \label{U}%
\end{align}
where $U_{c}\simeq\gamma^{3}/\Delta^{2}$ is the standard barrier. In our
regime $U\gg U_{c}$ the nucleation rate can be thus dramatically suppressed.

\section{Conclusions}

(1) Eq. (\ref{U}) shows clearly that homogeneous nucleation is prohibited in
nematic rubbers -- just as it is in martensites. Clearly, we need
heterogeneities to nucleate, and we do expect them in networks. (2) Even when
dealing with heterogeneous nucleation, it is reasonable to think that
platelets (or needles) close to the habit angle will be preferentially
generated. This would possibly be observed by electron microscopy or by force
microscopy on the outer surface. (3) From a practical point of view, it should
be beneficial to accelerate the commutation process -- we can think of at
least two ways: (a) adding colloidal platelets (clay ?) at the correct angle
(possibly orienting them by fields during the synthesis). (b) shearing the
sample with a shear plane near the habit angle.

\section{Acknowledgements}

We have benefited from exchanges with Y. Qu\'{e}r\'{e} and E. Rapha\"{e}l.

\end{document}